\documentclass[conference]{IEEEtran}

\usepackage{spconf,amsmath,graphicx}
\usepackage{multirow}
\usepackage{slashbox}
\usepackage{pstricks,cite}
\usepackage{pstricks}
\usepackage{epsfig,amsmath}

\title{Non-Linear Non-Stationary Heteroscedasticity Volatility for Tracking of Jump Processes}
\name{Seyyed Hamed Fouladi$^1$, Ehsan Hajiramzanali$^2$}
\address{$^1$Norwegian Institute of Science and Technology, Trondheim, Norway
\\
$^2$Texas A\&M University, College Station, TX, USA,
\\
\tt hamed.fouladi@ntnu.no, ehsanr@tamu.edu}

\begin{document}
\maketitle

\begin{abstract}
In this paper, we introduce a new jump process modeling
which involves a particular kind of  
non-Gaussian stochastic
processes with random jumps at random time points.
The main goal of this study is to provide an
accurate tracking technique based on non-linear non-stationary
heteroscedasticity (NNH) time series.
It is, in fact, difficult to track jump processes
regarding the fact that 
non-Gaussianity is an inherent feature in these processes.
The proposed NNH model is conditionally Gaussian whose conditional variance is time-varying.
Therefore, we use Kalman filter for state tracking.
We show analytically that
the proposed NNH model is superior to the
traditional methods. Furthermore, to validate the findings,
simulations are performed.
Finally, the comparison between the proposed method and other alternatives techniques has been made. 
\end{abstract}
\section{Introduction}
\label{sec:intro}

Non-linear and non-Gaussian state-space
models have attracted an increasing attention in recent years due to their significant
importance to 
a wide range of applications such as radar, biomedical signal and image processing, and wireless
communications \cite{hajiramezanali2018scalable,hajiramezanali2012isspa,delay,radar,hajiramezanali2018differential,hajiramezanali2018bayesian,radarconf,ali2011design}.
In these practical dynamic systems, 
the jump process is found to be helpful when abrupt changes occur within limited time intervals.
Therefore, the variation of state variables can be modeled by jump processes.
In \cite{SP}, it is mathematically proved that a time varying process
can be modeled by a heteroscedastic process properly. In \cite{ETRI,hajiramezanali2012ssp}, the generalized autoregressive conditional heteroskedasticity (GARCH) process was used as a residual error of state
model in the maneuvering target tracking which is one of the jumpy process applications.
The GARCH model outperformed
in comparison to the traditional models such as input estimation (IE), modified input estimation (MIE), and singer model.

These traditional models either fail to track jumpy points or result high errors in the flat or smooth parts of the jump processes.
In this paper, we solved this problem by proposing a new non-linear non-stationary heteroscedasticity (NNH) model.
The stochastic non-linear volatility leads the error covariance model to be time
varying. We consider time varying conditional covariance
for state equation which can describe both jumpy and non-jumpy time intervals.
The featured characteristic of the NNH model is the 
heterogeneity generating function (HGF) \cite{app}.
In our approach this function is considered as an exponential function.
This function is particularly useful for modeling of state space variance
when a jump process is to be tracked where a very large variance is observed 
at a jump point; meanwhile during flat intervals, low variance is experienced.
This new proposed model outperformed compared to other proposed models especially regarding steady state
time periods, because its error variance will be close to zero in these time slots.

The paper is organized as follows.
In Section 2, the jump processes are introduced and their
properties are explained.
The proposed model is presented in Section 3. 
Section 4 renders the modification of Kalman filter (KF) tracking algorithm to encompass our problem; i.e.,
modeling the stochastic jump process by the NNH time series for jump process tracking. 
The simulation results with various scenarios are given in Section 5.
Finally, the conclusion is given.

\section{Problem formulation}
\label{sec:format}

The purpose of this study is tracking of the jump process
The definition and properties of jump
process are discussed.

Following the literature \cite{IFAC}, compound Poisson processes is considered as a sub-class of the jump processes.
Compound Poisson processes are pure jump L\'{e}vy processes \cite{book}, with paths that are constant
apart from a finite number of jumps in finite time.

We consider the following linear discrete state space model formulation for modeling a jumpy system
\begin{eqnarray}
\label{model} x_{k} &=& A_{k} x_{k-1} + w_k, \\
\label{model2} y_k &=& H_k x_k + \nu_k,
\end{eqnarray}
where $x_{k}$ is a state vector at the $k$th instant,
$y_k$ is the measurement, $A_{k}$ is the transition matrix,
$H_k$ is the measurement matrix and $w_k$ and $\nu_k$ are modeling and
measurement noise with covariance matrices $B_k$ and $R_k$
respectively. Based on the pure modeling, the relation between current and
previous states is
\begin{equation}
\label{pure} x_k = \left\{ \begin{array}{ll}
x_{k-1},&\mbox{if no jump}\\
x_{k-1}+z,& \mbox{if jump occurs}
\end{array} \right.
\end{equation}
where the jump size $z$ is an arbitrary distribution.
In this formulation, we assume $z$ as a
Gaussian random variable which is a least informative distribution given the variance.
For (\ref{pure}), it is assumed that the sampling time $\Delta t = t_k - t_{k-1}$
is small enough relative to the reciprocal of the constant jump density ($1/\lambda$),
which is Poisson process parameter ($\lambda$). Therefore, more than one jump may happen
during this interval with
small probability which tends to zero.
Let us denote the number of Poisson points in one sampling interval
by $N(\Delta t)$, where $N(t)$ is a Poisson process with a constant jump density
$\lambda$, so
\begin{eqnarray}
\label{sampling1} P\{N(\Delta t) = 0\} &=& \exp(-\lambda \Delta t), \\
\label{sampling2} P\{N(\Delta t) = 1\} &=& (\lambda \Delta t)\exp(-\lambda \Delta t).
\end{eqnarray}
Since we have assumed that $(\lambda \Delta t) \ll 1$, then
\begin{equation}
\label{key} P\{N(\Delta t) = 0\} + P\{N(\Delta t) = 1\} \simeq 1.
\end{equation}
According to (\ref{key}), $P\{N(\Delta t) > 1\} \simeq 0$, which shows
more than one jump in sampling interval is completely rare.
Consequently, (\ref{model}), (\ref{pure}), and (\ref{key}), are simplified as
\begin{equation}
\label{final} x_k = x_{k-1} + \epsilon_k,
\end{equation}
where $ \epsilon_k $ is a non-Gaussian process noise which is given by
\begin{equation}
\label{noise} \epsilon_k = N(\Delta t) z_k + w_k.
\end{equation}
The distribution of this random variable
is approximated by solution of a SDE \cite{IFAC}.
In the next section, the mathematical formulation of the problem
of tracking jump process using the NNH volatility model
will be discussed.

\section{NNH for Jump process Modeling}

Autoregressive Conditional
Heteroscedasticity (ARCH) and GARCH models have widely been used to describe time varying conditional variances for many econometric
applications \cite{En, Bol}.
Although ARCH and GARCH models can capture abrupt changes well,
their performances deteriorate in the time interval with flat behavior.
The NNH model was proposed in \cite{NNH} in which it is considered
a time series with the conditional heteroscedasticities
captured by a non-linear function.
The NNH model provides a better result
in the flat time intervals while preserves a good performance at the jump points.
In the following, we establish the mathematical formulation of the model for jump process tracking.

We write the process noise model of (\ref{final}), $\epsilon_k$, as
\begin{equation}
\label{volatility} \epsilon_k = \sigma_k \xi_k,
\end{equation}
and let $I_k$ be a filtration, denoting the information available at time interval $[t_{k-1},t_k]$ and $\xi_k$ be an i.i.d. Gaussian random process with unit variance. Then,
$\epsilon_k$ is a zero mean process noise with conditional variance $\sigma_k^2$. 
\begin{equation}
\nonumber \mbox{\boldmath{E}}\{\epsilon_k|I_{k-1}\}=0, \qquad \mbox{\boldmath{E}}\{\epsilon_k^2|I_{k-1}\}=\sigma_k^2.
\end{equation}
The process noise model, $\epsilon_k$, is conditionally
heteroscedastic. 
Let the stochastic volatility, $\sigma_k^2$, be considered as
\begin{equation}
\label{sigma} \sigma_k^2 = f(h_k),
\end{equation}
for non-negative function, $f(h_k),$ and
\begin{equation}
\label{function} h_k = \rho h_{k-1} + \epsilon_k,
\end{equation}
where $h_k$ is an explanatory variable and by considering $\rho=1,$ $h_k$ becomes a non-stationary
integrated process.
Thus, explanatory variable, $h_k$, defined in (\ref{function}) becomes a non-stationary
integrated process. \textit{In this paper, we refer to $\sigma^2_k$ as NNH which constitutes our novel volatility model.}
The behavior of the proposed NNH model depends critically on the HGF, $f(h_k),$ in (\ref{sigma}).

\subsection{Estimation of HGF}

Let HGF, $f(h_k)$, in (\ref{sigma}) be specified in a parametric form as
\begin{equation}
\label{g} f(h)=g(h,\mbox{\boldmath $\theta$}),
\end{equation}
where $g$ is a known function and $\mbox{\boldmath $\theta$}$ is a set of unknown parameters.
Therefore, $\epsilon_k^2$ can be expressed as
\begin{equation}
\label{g2} \epsilon_k^2 = g(h_k,\mbox{\boldmath $\theta$}) + O_k,
\end{equation}
where
\begin{equation}
\label{O} O_k = f(h_k)(\xi_k^2-1).
\end{equation}
The nonlinear regression with integrated regressor such as the one in (\ref{g2}) has been
studied in \cite{NNH}. The HGF for jumpy behavior is subsequently fitted
to a parametric model $f(h) = g(h,\mbox{\boldmath $\theta$}) =\alpha \exp{(-{\beta}/{{h^2}})}$ using the weighted non-linear least squares
(WNLS) method introduced in \cite{NNH}. In what follows, we
simply assume that $g$ satisfies the conditions in \cite{NNH}.
Therefore, we model the stochastic non-linear volatility in terms of an exponential function as in
\begin{equation}
\label{func} g(h_k,\mbox{\boldmath $\theta$}) = \alpha \exp{\left(-{\beta}/{{h_k^2}}\right)}
\end{equation}
where $k$ is the time index, $\mbox{\boldmath $\theta$} = \{\alpha, \beta\}$, $\alpha > 0$ and $\beta>0$, are the unknown HGF parameters and $h_k$ are the explanatory variables.
As it is mentioned in (\ref{function}), $h_k = h_{k-1} + \epsilon_k$, therefore,
\begin{equation}
\label{alpha} g(h_k,\mbox{\boldmath $\theta$}) = \alpha\exp\left(\frac{-\beta}{(h_{k-1}+\epsilon_k)^2}\right),
\end{equation}
which according to  (\ref{final}), $\epsilon_k = x_k - x_{k-1}$, and using (\ref{sigma}), it is expressed as
\begin{equation}
\label{final2} \sigma_k^2 = \alpha\exp\left(\frac{-\beta}{{(h_{k-1}+(x_k-x_{k-1})})^2}\right), 
\end{equation}
where $\{\alpha,\beta\}$ are the unknown parameter to control the
behavior of HGF. The HGFs are considered as $e^h/(1+e^h)$, ${1+e^{-h^2}}$, $e^{-h^2}$ and $|h|$ in earlier research \cite{NNH}.
In the following section, we use the proposed new HGF in  (\ref{final2}) to model the volatility in Kalman filtering procedure in equation
(\ref{model}).
Subsequently, (\ref{model}) and (\ref{model2}) using (\ref{final2})
are explained in details to elaborate more our proposed NNH approach.

\section{Kalman Filtering Modification for NNH Modeling}
\label{sec:majhead}

\begin{figure}[t]
\centering
\centerline{\includegraphics[scale=.7]{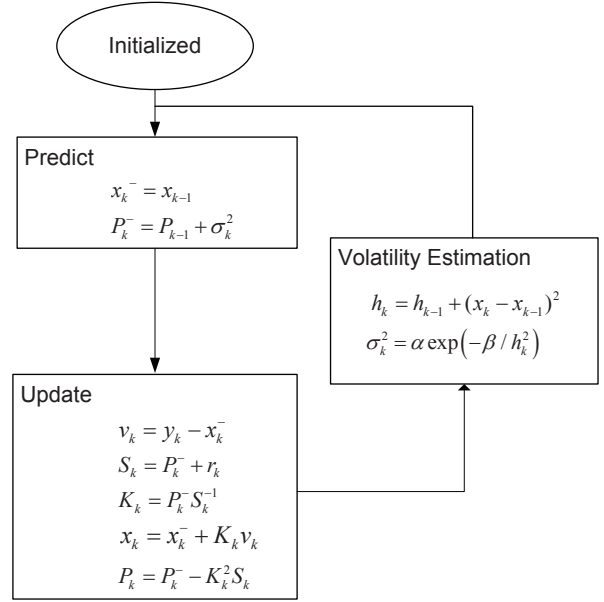}}
\caption{Functional view of the proposed NNH method.}
\label{fig}
\end{figure}

The state non-stationary equation given by (\ref{final}) has the structure for
which the linear Kalman filter approach fails because of non-Gaussianity of noise process $\epsilon_k$ in (\ref{final}).
But, an attractive feature of Kalman filtering is its ease of implementation which
we do not intend to forsake.
While the other filtering approaches; i.e., particle filters, can be used to estimate the target state
vector, but at very high  computational complexities.
Hence, we modify the Kalman filter procedure by adapting the
volatility of the process noise in each filtering step.
By this scheme, the non-Gaussian process noise is modeled by
time-varying variance Gaussian process. 
The obtained formulation is as follows
\begin{eqnarray}
\label{ali} x_k &=& x_{k-1}+\sigma_k\xi_k,\\ \nonumber
\sigma_k^2 &=& \alpha \exp{\left(\frac{-\beta}{{(h_{k-1}+(x_k-x_{k-1})})^2}\right)},\\
\label{ali3} y_k &=& x_k + \nu_k.
\end{eqnarray}
The volatility of motion equation (\ref{ali}), $\sigma^2_k$, is a stochastic process; therefore, its estimate is sought to
implement the modified Kalman filter. Hence, the performance of tracking obtained from the estimate of $x_k$ in (\ref{ali},\ref{ali3})
is sensitive to this
estimate. The proposed NNH estimator is dealt with the accurate estimation of $\sigma^2_k.$
We consider the parameter $\sigma^2_k$
as a random parameter and predict its sample at each step by (\ref{final2}), thus, our
uncertainties for state equation (\ref{ali}) in a highly jumpy process, is
compensated by adding a flexibility to determine $\sigma^2_k$ through (\ref{final2}).
The algorithm is initiated by a small value for $\sigma^2_1.$ Then, $\sigma^2_k$ is determined from  the update equation of Kalman
filtering in Fig.~\ref{fig} and (\ref{final2}). This modification
of variance during the algorithm, results in a better performance,
especially in a highly jumpy process. The flowchart in Fig.~\ref{fig} illustrates the functional structure of the proposed NNH method.

\section{SIMULATION AND RESULTS}
\label{sec:typestyle}

A set of scenarios is simulated to investigation of the performance of the proposed NNH
model in comparison with the traditional approaches.
These simulations involve an
ensemble of compound Poisson processes as the jump process. In the simulations, the observation interval
[$0,T$] is large enough compared to the sampling interval. The sampling time is
$\Delta t=0.1$ sec.
At first, we simulate and track the jump process with various parameters shown in Table \ref{TabComp}.
The variance of the measurement noise for our
simulation was selected as $r = 10$.
The variance of modeling noise, $Q = 5$ for the traditional Kalman filters.
The values of the proposed HGF parameters are assumed to be known as $\alpha = 50$ and $\beta=10$.
Simulations were run for different
parameters to demonstrate the robustness of the new approach versus the performances for Kalman filter and GARCH approaches.
For quantitative analysis, mean square error
(MSE) of tracking results are calculated. In Table~\ref{TabComp}, the average of MSEs over 200 trials are shown. As it can be seen in the table, the proposed model performs better than two other approaches.

\begin{table}
\caption{The Resulted MSE of Different Methods by 200 Monte Carlo simulations} \label{TabComp}
\begin{center}
\begin{tabular}{|c|c|c|c|c|}
\hline
 \multirow {1}{3.8cm}{\backslashbox [4cm][4cm]{Parameters}{Methods}} &{\centering KF}&{\centering GARCH}&{\centering Proposed } \\
  & & & NNH \\\hline
$\lambda=1, \mu=0, \sigma=10$& 6.78 & 4.14 &\bf 3.76 \\\hline
$\lambda=1, \mu=3, \sigma=10$& 5.96 & 4.06 &\bf 3.47 \\\hline
$\lambda=2, \mu=0, \sigma=8$&  7.13 & 5.32 &\bf 4.63 \\\hline
$\lambda=6, \mu=2, \sigma=20$&  8.2 & 5.27 &\bf 4.5 \\\hline
\end{tabular}
\end{center}
\end{table}
\begin{figure}[hb!] \label {fig2}
	\centering
	\begin{minipage}[ht!]{0.48\linewidth}
		\centering
		\centerline{\includegraphics[scale=.5]{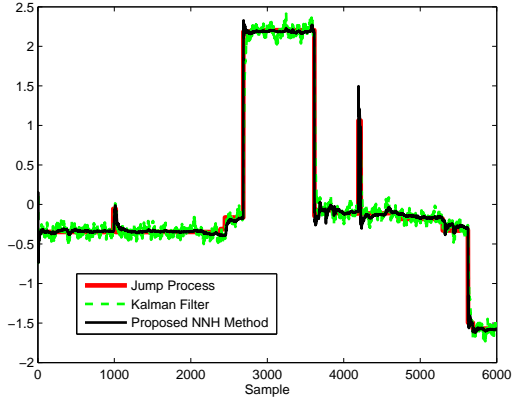}}
		\centerline{(a) The original and the estimated signal.}\medskip
	\end{minipage}
	
	\begin{minipage}[ht!]{0.48\linewidth}
		\centering
		\centerline{\includegraphics[scale=.5]{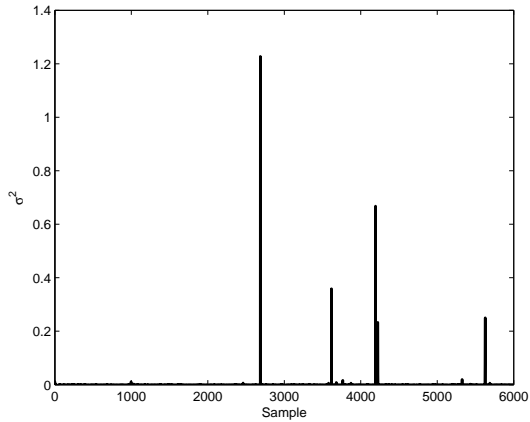}}
		\centerline{(b) The estimated volatility by the proposed NNH model}\medskip
	\end{minipage}
	
	\begin{minipage}[ht!]{0.48\linewidth}
		\centering
		\centerline{\includegraphics[scale=.5]{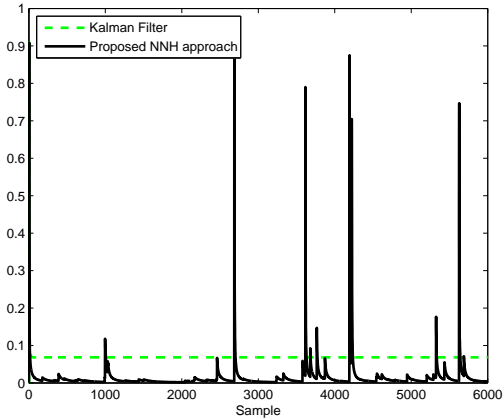}}
		\centerline{(c) The gains for Kalman filter and proposed NNH approach}\medskip
	\end{minipage}
	\caption{Simulation results: comparison of NNH and Kalman Filter}
	\label{fig:res}
\end{figure}
\begin{figure}[ht!]
	\center
	\centerline{\includegraphics[scale=.6]{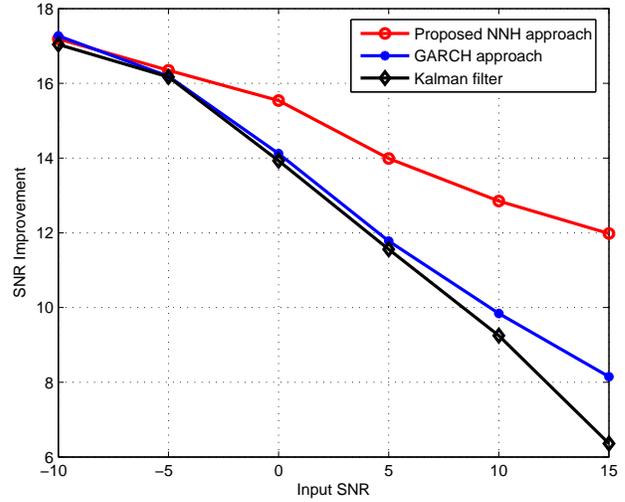}}
	\caption{SNR Improvements by different approaches in non-Gaussian noise.}
	\label{SNRImp}
\end{figure}

Moreover, Fig.\ref{fig:res}(a) illustrates the original signal and the estimated signals by the proposed method and the GARCH method. As it can be seen, the proposed model outperforms compared to the GARCH approach. In addition, the estimated volatilities of the proposed NNH model of this
jump process is shown in Fig.\ref{fig:res}(b).
In Fig.\ref{fig:res}(c), the gains $K_k$ in update step (in Fig.\ref{fig}) are represented for proposed NNH approach
and Kalman filter.
By looking at Fig.\ref{fig:res}, we observe $\epsilon^2$ increases when there is an abrupt change in the original signal (in jump condition).
On the other hand, for the time intervals with flat behavior, the estimated stochastic volatility is approximately
zero. However, Fig.\ref{fig:res}(c) shows that the variance of GARCH is near five in the steady state time slots.
Therefore, this feature of the proposed NNH method provides a smooth behavior in tracking
in comparison with the traditional methods during the flat behavior of signal. In other word, the proposed NNH model adapts
to the volatility of process noise.
Using this information, estimation filter has an acceptable
performance for abrupt change or jumps tracking.

Furthermore, in order to show the performance of our proposed tracking approach
in comparison with other methods, a jump process is generated
and is contaminated by an additive non-Gaussian noise with
different powers. We generate the non-Gaussian observation noise in equation (\ref{ali3}) by,
\begin{equation}
\nu_k=\kappa_k \eta_k
\end{equation}
where $\kappa_k$ is i.i.d. discrete uniform random process with values $\{-1,1\}$ and
$\eta_k$ is i.i.d. exponential random process.
We represent signal to noise ratio (SNR) Improvements by NNH method, GARCH approach and Kalman filter in Fig.\ref{SNRImp} in terms of input SNR.
The SNR Improvement is calculated as: {\it SNR Improvement (dB)= Output SNR (dB) - Input SNR (dB)}.
It can be seen that the proposed method
outperforms other approaches in non-Gaussian noise.

\section{Conclusion}
\label{sec:prior}
In this paper, we focused on tracking of jump processes. The properties of a jump process make it an
appropriate model for modeling signals and state variations in dynamic
systems with abrupt changes.
Since Kalman filter exploits an Gaussian model, it fails to track jump processes with non-Gaussian distribution.
Although GARCH process is a non-Gaussian model, this model has a high variance at non-jumpy points which leads to a poor tracking performance. The proposed approach uses NNH model with a nonlinear function estimating conditional variance over time. This gives the capability to have high variance values at jumpy points and very low variance values in other time intervals. This property is served to track sudden changes and also results a smooth estimated signal in non-jumpy parts. 
The simulation results showed that the traditional Kalman filter approach did not succeed under the circumstances as expected. 
We extended the Kalman filter by exploiting NNH model. By modifying the Kalman filter, the constant conditional variance of the process noise
is adapted to the stochastic volatility of the process noise by applying a non-linear function on
explanatory variable. The result of this model
described the jump process at the jump points and the time intervals with flat behavior well. Simulation results
illustrated that the proposed NNH method outperformed in comparison with the traditional approaches in both Gaussian and non-Gaussian noise scenarios.

\bibliographystyle{IEEEbib}
\bibliography{refs}

\end{document}